\input{aipcheck}
\documentclass[
    ,final            
  ]
  {aipproc}

\layoutstyle{6x9}

\begin{document}

\title{Topological spectrum of classical configurations}

\classification{
                \texttt{02.40.Hw,03.70.+k}}

\author{Francisco Nettel}{
  address={Instituto de Ciencias Nucleares \\ 
  Universidad Nacional Aut\'onoma de M\'exico \\
  A.P. 70-543, M\'exico D.F. 04510 \\
  M\'exico }
}
\author{Hernando Quevedo}
{}
\begin{abstract}
For any classical field configuration or mechanical system with a finite number of
degrees of freedom we introduce the concept of topological spectrum. It is based upon
the assumption that for any classical configuration there exists a principle 
fiber bundle that contains all the physical and geometric information of the configuration.
The topological spectrum follows from the investigation of the corresponding 
topological invariants. Examples are given which illustrate the procedure and the
significance of the topological spectrum as a discretization relationship among 
the parameters that 
determine the physical meaning of classical configurations.
\end{abstract}

\maketitle


\section{Introduction}

Canonical quantization is one of the keystones of modern physics, 
especially due to the fact that its application to gauge field theories 
has lead to the development of the standard model of elementary particles, whose 
predictions have been corroborated in the laboratory with spectacular accuracy.
From a pure theoretical and mathematical point of view, nevertheless, canonical 
quantization presents severe problems. Among them one can mention 
the assumption of the existence of a Hilbert space that indeed exists only 
in very particular cases, 
the existence of operators that in the case of field configurations are 
not always well-defined, the problem of the classical limit, 
the non-uniqueness of the quantum evolution, the divergencies that require 
the application of different regularization and renormalization procedures, and many 
others  \cite{wald,carlip}. In the case of gravity, the situation is such that even 
if one would be able to solve all the technical problems of canonical quantization, 
one still would be confronted with the problem of time that confronts us with our
conceptual understanding of space and time (see, for instance, \cite{macquev} for a 
recent review).  

In view of this situation, during the past 60 years many authors have been trying to 
formulate alternative quantization procedures (see, for instance, \cite{carlip} for an
introductory review). Unfortunately, none of these methods have reached the tremendous 
experimental success of canonical quantization. 
Recently, we started an alternative approach based upon the topological 
properties that can be extracted from a principal fiber bundle, associated 
to any classical physical configuration \cite{patquev}. We called this method  
topological quantization. It has been used previously
in the context of diverse monopole and instanton configurations 
\cite{frankel}.  We have shown that when applied to certain gravitational fields,
topological quantization leads to a set of discretization conditions on the
parameters that determine the field. Moreover, a preliminary study \cite{netquev} 
has shown that 
the method can be applied to mechanical systems with a finite number of degrees of 
freedom. In particular, it was possible to find discretization conditions that 
are equivalent to the spectrum following from the canonical quantization of
the harmonic oscillator. 

The aim of the present work is to begin a more strict 
formulation of the method of topological quantization. 
We introduce the concept of classical configuration, considering the geometric
structures needed in topological quantization, and 
construct principal fiber bundles that can be associated to any given classical 
configuration. The concept of  topological spectrum is explained by 
using topological invariants.

\section{Classical configurations}
\label{sec:cc}

Although the idea of classical configurations is very intuitive and no exact 
definition is usually necessary in physics, for the purposes of topological 
quantization we need a more mathematical approach. 
We introduce the concept of classical configuration as any classical 
(non quantum) physical system to which we can associate a unique geometric 
structure consisting of a differential manifold with a connection. Uniqueness 
should be understood here in a flexible manner as far as these two geometric 
objects are sufficient to distinguish classical configurations from each other. 
So, for instance, two differential manifolds which are related by an isomorphism 
with the same connection describe in our formalism the same classical configuration.

To be more specific let us consider examples from field theory. In Yang-Mills 
gauge theories, a classical configuration is a 
solution of the corresponding field 
equations. The differential manifold is the Minkowski spacetime and the connection 
$A$ is a differential 1-form with values on the algebra ${\bf g}$ 
of the gauge group $G$ which can be $U(1)$, $SU(2)$ or $SU(3)$ for the known 
gauge interactions of nature. Any solution of the field equations can be represented,
up to a gauge transformation, by means of a connection $A$ which generates the 
gauge curvature $F= dA$ in the Abelian case or $F=dA + A\wedge A $ in the non-Abelian 
case. Although the physical information of the configuration is invariantly contained
only in the gauge curvature, we will use the connection because, as we will see in
the next section, its gauge freedom is 
an important component for the construction of the underlying geometric structure 
of topological quantization.  A classical configuration in gauge field theories will 
then be denoted by $(M_\eta , A)$.
Notice that the corresponding solution of the Yang-Mills equations does not necessarily
need to be exact. Approximate solutions are also allowed, as far as they can be associated 
with gauge connections.

Gravitational fields are further examples of classical configurations. Let $g$ denotes
an exact or approximate solution of Einstein's equations in vacuum. According to 
general relativity, the Riemannian manifold $M_g$ with metric $g$ 
is the geometric object that contains all the relevant information about the 
gravitational field. In this case, a classical configuration will be denoted by
$(M_g, \omega)$ where $\omega$ is the spin connection following from $g$. In fact, 
if we introduce a local
 orthonormal vierbein $\theta^a$ by means of 
 $g=\eta_{ab}\theta^a \otimes \theta^b$, where $\theta^a = e^a_{\ \mu} dx^\mu$ and $x^\mu$ are spacetime coordinates,
then the spin connection is determined by $d \theta^a = -\omega^a_{\ b} \wedge \theta^b$
and takes values on the algebra of the Lorentz group $SO(1,3)$. The corresponding
curvature 2-form 
$\Omega^a_{\ b} = d\omega^a_{\ b} + \omega^a_{\ c} \wedge \omega^c_{ \ b}$ 
is again the basic
geometric object from which physical properties of the corresponding gravitational
field can be extracted. Notice that we have chosen a local description of the
gravitational field in terms of the orthonormal differential 
frame $\theta^a$, instead of the usual tensorial approach with spacetime coordinates 
$x^\mu$ for which we would have a classical configuration as the pair $(M_g, \Gamma)$,
where $\Gamma$ is the Levi-Civita connection. The advantage of the local approach 
in terms of differential forms is that we reduce the diffeomorphism invariance
of the metric approach to the local invariance of the Lorentz group. The Lorentz invariance
and the corresponding spin connection are clearly easier to handle from a geometric 
point of view. In contrast to gauge theories, in general relativity the metric $g$ 
completely determines the Levi-Civita connection $\Gamma$ (or the spin connection 
$\omega$) so that for a classical configuration we only need to know $g$. Nevertheless,
we use the notation $(M_g,\omega)$ to emphasize the fact that it is possible to consider 
more general theories of gravity in which the connection is not compatible with the
metric.

Mechanical systems with only a finite number of degrees of freedom can also
be considered as classical configurations. Recall that in classical mechanics a system
with $k$ degrees of freedom is given by a Lagrangian of the form 
$L=(1/2)g_{\alpha\beta}\dot q ^\alpha
\dot q^\beta - V(q)$ ($\alpha,\beta=1,...,k$). We limit ourselves to conservative systems 
in which the Hamiltonian is a constant of motion which coincides with the total 
energy $E$ of the system.  It turns out \cite{netquev} that a way
to differentiate mechanical systems from each other is through the Jacobi metric 
$h=2(E-V)g_{\alpha\beta} dq^\alpha dq^\beta$, 
which is also used in Maupertuis' formulation 
of classical mechanics \cite{arnold}. In fact, if one introduces Cartesian coordinates,
the metric $g_{\alpha\beta}$ 
becomes proportional to the Euclidean metric $\delta_{\alpha\beta}$, and the
conformal factor $(E-V)$ will contain in the potential all the information about the 
physical system. The properties of mechanical systems are invariant with respect to 
Galilean transformations. If we introduce a local orthonormal frame $\theta^i$ 
such that $h=\delta_{ij}\theta^i\otimes \theta^j$, the invariance becomes reduced
to that of the rotation group $SO(k)$. The differential 1-forms $\theta^i$ generate
a rotation connection $\omega^i_{\ j}$ as described above in the case of gravitational fields. Consequently, a mechanical system can be interpreted as a classical configuration
with the pair $(M_h, \omega)$, where $M_h$ is a $k$-dimensional conformally flat manifold
with metric $h$. Notice that we do not need to know the solutions of the Euler-Lagrange
equations in order to study mechanical systems as classical configurations. This is
a property that could be of advantage, especially when investigating mechanical systems
with potentials $V(q)$ which do not allow an analytical integration of the motion 
equations. 

The above examples show that the class of classical configurations includes all possible 
conservative mechanical systems, all vacuum gravitational fields, and all solutions 
of the Yang-Mills field equations. In the case of field configurations, one can now
consider any combination of the gravitational field with different types of 
gauge matter to generate new classical configurations. One could say that any 
classical solution of the field equations for the four fields observed in nature, 
or an arbitrary combination
of them, can be considered 
geometrically as a classical configuration. 

It is worth mentioning that the scalar
field in its standard Lagrangian description is not a classical configuration 
as defined above. 
There
is no natural connection that could be associated to the scalar field, although 
it possesses the natural differential manifold structure of the Minkowski spacetime. 
In the context of topological quantization the scalar field requires a special treatment  
which will be presented elsewhere.   

\section{Topological spectrum}

According to our definition, a classical configuration is characterized by the pair $(M,\omega)$ consisting of a differential manifold $M$ and a connection 1-form  $\omega$.
Let us suppose that the physical content of $(M,\omega)$ is invariant with respect to
transformations of a group $G$. In the case of mechanical systems, $G$ is clearly the
Galilean group which reduces to the rotation group $SO(k)$ when a local orthonormal vielbein is used in the description. In a similar manner, gravitational fields are in general locally 
invariant with respect to transformations of the Lorentz group $SO(1,3)$ acting on the 
differential 1-forms $\theta^a$. Gauge field configurations are invariant with respect 
to transformations of the corresponding gauge group, when acting on the connection $A$, 
and of the Lorentz group, when acting on the underlying Minkowski metric $\eta$. 
As we will see below only the gauge invariance is not trivial in this case. 

We now use the triplet $(M,\omega,G)$ to construct a principal fiber bundle $P$ with a 
connection ${\tilde \omega}$ in the following way. Let $M$ be the base space of $P$. To
each point of $M$ we attach the elements of the group $G$ as the typical fiber which 
is isomorphic to the structure group of $P$. Furthermore, let $\sigma_i$ be 
the local section over an open subset $U_i\subset M$ which follows in a standard manner
\cite{naber}  
from the local trivialization of $P$ over $U_i\times G$. The connection on $P$ 
is introduced by means of the condition $\sigma_i^* \tilde \omega = \omega_i$ where
$\omega_i$ is the connection $\omega$ of the base space $M$ evaluated on $U_i$, 
and $\sigma_i^*$ is the pullback of $\sigma_i$. In \cite{patquev} it was shown
that these elements are sufficient to construct all the constituents of 
a principal bundle $P$ with connection $\tilde \omega$. Moreover, for the triplet
$[M_g,\omega,SO(1,3)]$, which corresponds to a specific 
vacuum solution of Einstein's equations,  one can prove the uniqueness of $P$.
One could expect that an analogous proof could be formulated for gauge fields. 
In the case of mechanical systems a detailed proof is beyond the scope of the
present work and will be presented elsewhere. 

It is worth 
noticing that the old idea of additional dimensions to describe the physical 
behavior of fields is incorporated in a natural way in our
construction. For instance, for gravitational fields we have that dim($P$) = 
dim($M_g)+$dim$[SO(1,3)] = 10$. A Yang-Mills field theory with gauge group $SU(k)$ on 
the Minkowski spacetime $M_\eta$
will be described on a principal fiber bundle of dimension $4 + k^2 -1$. Classical
configurations of the standard
model of elementary particles with gauge group $U(1)\times SU(2)\times SU(3)$
will be described on a 16-dimensional principal bundle. 
Mechanical systems with $k$ degrees of 
freedom are characterized by a $k$-dimensional base space $M_h$ with metric $h_{ij}$ so
that the dimension of the corresponding $P$ is given by 
dim$M_h$ + dim$[SO(k)]= k + k(k-1)/2$. 
Although those additional dimensions
have very important physical consequences, in the sense that they are related to
physical symmetries of the system, they cannot be observed directly on the spacetime
which is the base space of $P$. In other words, our additional dimensions are important
for the geometric construction of $P$, but they manifest themselves on the spacetime $M$
only through conservation laws which can be associated to symmetries of 
the system.

Since to any classical configuration $(M,G)$ we can associate a principal fiber bundle $P$, we can use the invariant properties of $P$ to characterize each configuration. A 
characteristic class $C(P)$ is a topological invariant of the bundle $P$; in addition,
the integral $\int C(P)$ over the base manifold $M$ (or over a compact cycle of $M$) is 
also an invariant. The remarkable result is that $C(P)$ can always be normalized in 
such a way that $\int C(P) = n$, where $n$ is an integer \cite{damas}. However, for this 
to be true, it is necessary that the integral could be computed, i.e., $C(P)$ must be
a differential form. Fortunately, for the cases of interest in this work, $C(P)$ can 
be expressed in terms of the curvature 2-form $\Omega$. The explicit form of $C(P)$ 
depends on the structure group $G$ of $P$ (for more details see, for instance,
\cite{nashsen}). If $G=O(k)$, one has the Pontrjagin class 
$p(P)$; if $G=SO(k)$, one has the Pontrjagin class $p(P)$  and the Euler class $e(P)$ 
which is non-zero only when $k$ is even; finally, for $G=U(k)$ one obtains the Chern
class $c(P)$. In all these cases, the integral $\int C(P)$ can be calculated 
explicitly and the result is a function $f(p_1,..., p_s)$ of the parameters that
enter the metric of the base space $M$ and, consequently, the curvature $\Omega$. 
So, as a result of the integration of the characteristic classes of $P$ we 
obtain a relationship of the form 
\begin{equation}
f(p_1,..., p_s)=n \ . 
\end{equation}
This is what we call the 
topological spectrum of the underlying classical configuration. It represents a 
discretization of the parameters $p_1, p_2,..., p_s$ 
which determine the physical significance 
of the base space $M$. The following examples illustrate the result of applying 
this procedure to specific classical configurations.

Consider a mechanical system consisting of two harmonic oscillators of mass $m$
\begin{equation} 
L = \frac{1}{2}m\left[ (\dot q ^1)^2 +  (\dot q ^2)^2 \right ]
- \frac{1}{2}\left[ k_1 (q ^1) ^2 + k_2( q ^2) ^2\right]\ .
\end{equation} 
The second oscillator is needed only to avoid the degeneracy the Jacobi metric
for a single oscillator. When written in the local zweibein $\theta^1 =\sqrt{2m(E-V)} dq^1$
and $\theta^2 =\sqrt{2m(E-V)} dq^2$, this classical configuration
is invariant with respect to the group $SO(2)$. After the construction of the 
corresponding principal bundle $P$ as described above, we find that the relevant characteristic class is the Euler class $e(P) \propto \Omega_{12}\theta^1\wedge \theta^2$.
Its integration in the limiting case of a single oscillator ($k_2=0, \ k_1=k)$ yields 
\cite{netquev}
\begin{equation}
\frac{kq_0}{kq_0^ 2 - 2 E} = n \ ,
\end{equation}
where $q_0$ is a parameter related to the turning point of the oscillator. This is the
topological spectrum of the harmonic oscillator. The constant $q_0$ can be chosen such
that one obtains the canonical spectrum from the topological one.

Consider the gravitational field of the Kerr-Newman black hole which is completely
determined by the mass $m$, specific angular momentum $a$, and charge $e$. It is
a solution of the Einstein-Maxwell equations. The base space of $P$
corresponds to the spacetime $M_g$ for this black hole. For the structure 
group we have two immediate possibilities: either $G=SO(1,3)$ or $G=U(1)$. For the 
sake of simplicity, let us consider the 5-dimensional principal bundle $P$ 
with structure group $U(1)$. The connection $A$ is a $u(1)$-connection from which 
the gauge curvature $F$ can be computed. The Chern class $c(P)$ is in this case the
relevant characteristic class whose integration over the base space results in the 
topological spectrum \cite{patquev}
\begin{equation}
\frac{2e^3 \sqrt{m^2-e^2-a^2}}{r_0(e^4 + 4 m^2 a^2)} = n \ , 
\end{equation}
where $r_0$ is an integration constant. In the case of vanishing angular momentum, $a=0$,
this topological spectrum can be rewritten in terms of the horizon area ${\cal A}$ of the 
Reissner-Nordstr\"om black hole as
\begin{equation}
{\cal A} = 4\pi e^2 {\cal A}_0 \left[\frac{n}{2}+\sqrt{1+\frac{n^2}{4}}\right]^2\ ,
\end{equation}
where ${\cal A}_0$ is a constant. We see that in this case the topological spectrum 
corresponds to a discretization of the horizon area. Further examples of gravitational 
classical configurations are given in \cite{patquev}.

In the case of the principal bundle for classical gauge configurations 
we also have two different possibilities for the structure group: 
the Lorentz group which 
represents the invariance of the background Minkowski spacetime, and the gauge group 
which corresponds to the gauge invariance of the field. Since the spin connection of the background metric is flat, the corresponding characteristic class vanishes identically
and no topological spectrum is obtained. However, if we take the gauge group as the 
structure group of $P$ a topological spectrum emerges. Probably, the simplest example of 
a topological spectrum for a gauge theory is the one which follows from analyzing
the field of an electric charge \cite{damas}. The resulting principal bundle $P$ is
a $U(1)-$bundle whose Chern number implies the topological 
quantization of the electric charge.

\section{Conclusions}
\label{sec:con}

In the present work, we presented a geometric definition of classical configurations which 
leads to the natural introduction of a principal fiber bundle with a connection. 
It can be applied to solutions of differential
field equations, such as the gauge fields and the gravitational field, and mechanical 
systems with a finite number of degrees of freedom. We showed that the study  
of the corresponding characteristic classes of the principal bundle leads to a topological
spectrum, represented by a discretization relationship among the parameters which 
determine the physical significance of the underlying classical configuration.





\end{document}